\documentclass[sigconf]{acmart}

\AtBeginDocument{%
  \providecommand\BibTeX{{%
    \normalfont B\kern-0.5em{\scshape i\kern-0.25em b}\kern-0.8em\TeX}}}

\copyrightyear{2019}
\acmYear{2019}
\setcopyright{acmlicensed}
\acmConference[HYPERTEXT '19]{HYPERTEXT '19: ACM Hypertext }{September 17--20, 2019}{Germany}
\acmBooktitle{HYPERTEXT '19: ACM Hypertext, September 17--20, 2019, Germany}
\acmPrice{15.00}
\acmDOI{10.1145/1122445.1122456}
\acmISBN{978-1-4503-9999-9/18/06}

\begin{document}

%
\title{Audience and  Streamer Participation at Scale on Twitch}

%
\author{Claudia Flores-Saviaga}
\email{cif0001@mix.wvu.edu}
\affiliation{
  \institution{West Virginia University}
  \city{Morgantown}
  \state{West Virginia}
  \country{USA}
}

\author{Jessica Hammer}
\affiliation{
  \institution{Carnegie Mellon University}
  \city{Pittsburgh}
  \state{Pennsylvania}
  \country{USA}
}

\author{Juan Pablo Flores}
\affiliation{
  \institution{Universidad Autonoma de Mexico (UNAM)}
  \city{CDMX}
  \country{Mexico}
}

\author{Joseph Seering}
\affiliation{
  \institution{Carnegie Mellon University}
  \city{Pittsburgh}
  \state{Pennsylvania}
  \country{USA}
}

\author{Stuart Reeves}
\affiliation{
  \institution{University of Nottingham}
  \city{Nottingham}
  \country{UK}
}

\author{Saiph Savage}
\affiliation{
  \institution{West Virginia University}
  \city{Morgantown}
  \state{West Virginia}
  \country{USA}
}



 




%
\renewcommand{\shortauthors}{Flores-Saviaga, et al.}

%
\begin{abstract}
Large-scale streaming platforms such as Twitch are becoming increasingly popular, but detailed audience-streamer interaction dynamics remain unexplored at scale. In this paper, we perform a mixed methods study on a dataset with over 12 million audience chat messages and 45 hours of streamed video to understand audience participation and streamer performance on Twitch. We uncover five types of streams based on size and audience participation styles: Clique Streams, small streams with close streamer-audience interactions; Rising Streamers, mid-range streams using custom technology and moderators to formalize their communities; Chatterboxes, mid-range streams with established conversational dynamics; Spotlight Streamers, large streams that engage large numbers of viewers while still retaining a sense of community; and Professionals, massive streams with the stadium-style audiences. We discuss challenges and opportunities emerging for streamers and audiences from each style and conclude by providing data-backed design implications that empower streamers, audiences, live streaming platforms, and game designers.
\end{abstract}

%
%

%

%

%
\maketitle

\section{Introduction}
Live streaming platforms, such as Twitch and  Youtube Gaming, Periscope, and Hitbox, have become considerably popular in recent years \cite{hamilton2014streaming,twitchPopular2,twitchPopular}. Streamers on these platforms each have a channel where they generally live stream themselves engaging in various entertaining activities from engaging at an event, creating art or music, or playing video games while interacting via chat messages with an audience who can be globally distributed \cite{twitchPopular}. This dynamic has created new types of experiences and interactions that we have yet to understand. 

We focus in this paper on Twitch both because of its scale and popularity and because of its wide distribution of types of communities; while Twitch is still primarily a game streaming platform, it has branched out significantly in recent years to encourage creative streaming and In-Real-Life (IRL) streaming. There is vast HCI research on audience and spectator engagement, including within gaming  \cite{cheung2011starcraft}, but our understanding of how audiences and streamers collectively participate in live game streaming platforms is still limited \cite{cheung2011starcraft, Ford2017Chat}. Most existing studies have focused on small scale audience-gamer interactions, for instance how small groups of gamers and audiences collaborate at home or in arcades \cite{tekin2017ways,Harper:2013:MGS:2441776.2441797,lin2011role}. But, very recent late breaking work \cite{Ford2017Chat} has showed the significance of this limitation, identifying critical relationships between audience size and audience behavior. 

While tens of millions of audience members participate on  Twitch \cite{audienciasChicas}; many streamers struggle to engage with their audiences and grow their communities in ways they'd like. This has lead to situations where individual streamers have limited viewership or struggle to engage productively with certain types of viewers \cite{TwitchRetro:2016,audienciasChicas2,audienciasChicas3}. The desire to entertain and drive audiences to participate arises not only because it generates better user experiences \cite{webster1997audience}, but also because Twitch's financial model rewards streamers with highly participatory audiences \cite{twitchDinero}. 

Seering et al., propose the framework of ``audience participation'', specifically within game design, to explain the dynamic of participation on Twitch \cite{seering2017APG}. By better understanding how the performance of streamers interacts with \textit{audience participants}, we can improve the design of live streaming platforms by bettering  user experience, recruiting more viewers, and increasing as a result the financial feasibility of pursuing streaming as a career path.  Analyzing audience-streamer participation at different scales also facilitates the creation of tools for a wider range of individuals and audiences. For instance, it enables the design of platforms tailored not only for professional streamers who have massive audiences, but also for novice streamers who may use the platform as a part-time activity or hobby and hence have more sporadic audiences who participate with them. 

To unpack this critical aspect of streamer-audience participation, we take a case-study approach, examining Twitch, the most popular and significant live game streaming platform \cite{twitchPopular}. Started in 2011, Twitch is now the world's leading social platform and community for video game culture \cite{twitchPopular3}. Twitch acts as a medium for audience members to watch highly skilled gamers play, discover new video game content, and engage and participate with streamers or other audience members. Twitch enables us to investigate audience and streamer participation in game spaces at a range of \textit{scales}. 

We conduct a large scale data analysis on Twitch to understand audience \textit{participation} in different scales, as well as the techniques streamers use to drive participation of different sized audiences. This analysis can help us to understand audience needs around spectatorship and participation, how differently-sized audiences are similar and different, and the effectiveness of streamer techniques when used within different contexts. Specifically, we identified the following research questions as fundamental to understanding audience participation in Twitch streams:

\begin{itemize}
  \item \textit {RQ 1.} How does audience participation on Twitch vary across different sized audiences?
  \item \textit{ RQ 2.} What type of techniques do streamers use to drive audience participation and how do these techniques vary?
\end{itemize}

We explored these research questions using one month of data from 130 randomly selected Twitch streams, totaling 2,700 minutes of video of streamers in action and 12,150,866 audience chat messages. We used cluster analysis to categorize streams, tracked how audiences participated in those streams throughout time, and applied sentiment analysis to understand audience participation in chat. We also used qualitative techniques to examine more closely the techniques that a selection of streamers from each of the clusters used to drive interactions with audiences. In particular, we build off audience participation frameworks \cite{levinson1988putting} to make concrete notions of how streamers ``perform'' for an audience, i.e., the type of techniques they adopt. 

Through our analysis we demonstrate that audiences and streamers participate differently within different sized groups. In specific cases, streamers struggled in balancing the entertainment of their audience, maintaining a certain self-presentation, and playing their game. We address this challenge by discussing design implications of our findings and provide data-backed recommendations for empowering streamers, audiences and designers. We discuss how our results can inform designers of these live streaming platforms to improve the experience of both viewers and streamers.

\section{Related Work}

Livestreaming platforms provide an opportunity for audiences to attend and participate in live events \cite{hamilton2014streaming,deng2016}. In the category of gaming, audience members join online to see and hear streamers as they play. An audio/video feed of the streamer's game---often superimposed with the streamer on webcam---is paired with a public text chat \cite{hamilton2014streaming,lessel2017}. The audience uses the chat to communicate with the streamer and each other. Streamers follow along and converse audibly with the chat, as they play \cite{hamilton2014streaming}. Together, livestreaming interfaces facilitate a shared community space and discourse \cite{bingham2017,lesselmauderer2016}. 

\subsection{Streamers on Twitch}

In the Twitch context, streamers are content creators and broadcasters. Beyond just distributing feeds of video games, streamers create dynamic live content based on their personality and skill. The resulting product is the performance that seeks to engage viewers \cite{bingham2017}.

Streamers can select from a variety of game content to deliver to their audiences. When broken down into categories of gameplay, the most common include e-sports, "speedrunners" and "Let's Play" \cite{smith2013}. Much of the research on live-streaming communities focuses on esports \cite{bingham2017,sjoblom2017}, which can range from individual amateur gameplay to large-scale tournaments viewed by millions of fans \cite{bingham2017,cheung2011starcraft,deng2016}. Speedrunning streamers demonstrate a high level of gameplay skill, and attempt to play through games at high speeds. Game ``bugs'', ``glitches'', and ``exploits'' are exploited to facilitate the game's speedy completion \cite{bingham2017,smith2013}. Let's Play enables the audience to play games by converting typed chat commands into gameplay commands \cite{lesselmauderer2016,smith2013}.

Using third-party tools, streamers provide additional features and enhancements to curate the stream content and game experience for their viewers. Financial motivations exist for streamers to develop audiences that are socially driven as they can add to the stream's ability to generate revenue \cite{beyer1997,savage2015}.

\subsection{Audiences on Twitch}

The success of Twitch demonstrates that a great demand exists for video game spectatorship \cite{sjoblom2016}. However, viewer motivations for attending live-streaming events are still being discovered. Hamilton describes audiences' potential motivations for joining the channel of a specific Twitch streamer as a split between possessing a desire for the particular content of the stream (including here the streamers' personality, skill and game being played), as well as participating in the community built around the specific channel (i.e., interacting with the other audience members of the channel) \cite{hamilton2014streaming}. Others see desires for entertainment, educational motives \cite{drucker2002spectator}, desire to improve their own gameplay, \cite{cheung2011starcraft,sjoblom2017}, and a sense of being in shared attendance when game moment highlights occur \cite{hamilton2014streaming,robinson2016}.

Although stream content types vary widely, studies indicate viewers gravitate to the most popular streams \cite{harpstead2015}, and overall watch a limited number of games \cite{deng2016}. While the number of Twitch viewers is steadily growing \cite{TwitchRetro:2015,TwitchRetro:2016}, the number of streamers and the average number of hours streamed is growing even faster \cite{Ford2017Chat,TwitchRetro:2015,TwitchRetro:2016}. Additionally, the Twitch interface promotes streams with large audiences, creating a feedback loop in which large streams can grow faster than small ones \cite{twitch}. Taken together, these factors produce a heavy-tailed distribution of viewers across streams, with a small number of extremely large streams and an extremely long tail of streams with few or no viewers \cite{bingham2017}.

Stream viewers change their behaviors depending on the size and activity of other community members. For example, in streams with increased numbers of active users in chat, messages are shorter and utilize more emoticons \cite{olejniczak2015}. In larger streams, audiences react to the high quantity of participation and fast scrolling chats by adopting various coherence techniques and writing practices \cite{Ford2017Chat}. These findings reflect features of the Twitch chat interface, such as how long messages stay on a viewer's screen. However, they may also reflect more general features of how audience size affects communication. When speaking to a smaller audience, messages tend to be more defined, concrete and display a greater attention to needs and situations of others. As audiences grow, messages become more abstracted with a greater focus on self-presentation, expressing aspects from one's experience often in a positive light \cite{barasch2014broadcasting}. 

\subsection{Participation and Retention on Twitch} 

Twitch audiences participate in the live-streaming environment by use of the chat interface. 
Chat participation is largely social \cite{hamilton2014streaming}, with frequent use of community-specific neologisms and emoticons \cite{olejniczak2015}. 
Let's Play and Audience Participation Games are notable exceptions, in which audiences can use the chat to participate in gameplay (e.g. choice chamber \cite{choicechamber}, beatstep cowboys \cite{beatstep}). While still uncommon, these categories of games offer interactivity that allows for direct game and streamer impact \cite{seering2017APG,vosmeer2016}.

By participating socially in chat, audiences have the ability to affect streamers in informal ways, as in influencing their selection of games and use of in-stream tools \cite{lessel2017}. Audience members who engage with a stream's chat typically adopt community-specific language and means of interaction \cite{olejniczak2015}. Additionally, as user behavior patterns are likely to be imitated, participation in a live-streamed community can help shape positive or negative behavior within the stream. Communities with active participants have the potential to create valuable connections and alliances among its members \cite{seeringkraut2017}.


Once acquired, persistence among Twitch audiences is high, with a third of streaming sessions lasting 60-120 minutes \cite{zhang2016,nascimento2014}. Streamers also use external tools and service like Twitter and Discord to communicate with audiences during off-line hours to remind them to join the stream and encourage viewer retention \cite{bingham2017}. However, little is known about how viewers are retained across multiple sessions or what factors influence them to persist.


\section{Methodology}

\subsection{Data Collection}
In order to understand audience participation and streamer performance on Twitch at different scales, we used Twitch's API to scrape data from all English-language streams that allowed free, public participation (i.e., it was not necessary for people to pay to subscribe to participate) between April 10, 2017 and May 17th, 2017. Our scraping script collected stream-level metadata including the stream's name, data about the game currently being played, the title of the stream, the size of the audience and time-stamp. Using a bot that remained in the stream until it went offline, we collected all messages sent in the stream's associated IRC (Internet Relay Chat). This collected the text of the messages, the name of the stream where the messages were posted, time-stamp and user-name of the sender. Video stream data was also downloaded for all active streams. For the purpose of this paper, we analyzed a subset of these video streams. See Table \ref{table:Datasets} for details.

\begin{table}[htp]
  \centering \small
  \begin{tabular}{p{5cm} p{2cm}}
    \hline
        { Days Collecting Data}  & 44 \\
        { Number of Streamers (Twitch streams)}  & 226,658 \\
        { Minutes of Analyzed Video Stream}  & 2,700  \\
        { Number of Viewers Participating in Chat }  & 651,664  \\
        { Number of Chat Messages}  & 12,150,866  \\ \hline
  \end{tabular}
  \caption{Twitch Data Collection}
  \label{table:Datasets}
\end{table}
\vspace{-5mm}

\subsection{Methods}
We focus our analysis on understanding interactions between streamers and their audience at different \textit{scales}. For this purpose, we: (1) use standard clustering techniques to group and uncover the different audience sizes (scales) present in Twitch; (2) model the participation of audiences within each cluster; (3) conduct qualitative analysis on the video of streamers from each cluster to understand how streamers perform differently and similar for different sized audiences.

\subsection{Uncovering Types of Audience Scales on Twitch.}
Previous work had started to investigate audience dynamics on Twitch at different scales. However, such research used arbitrarily-defined categories for audience size. For instance, \cite{hamilton2014streaming} defined small audiences as involving less than 1,000 audiences members, while massive audiences had over 1,000 individuals. We instead use a quantitative clustering approach to uncover different audience scales present on Twitch.

For each stream, we calculated their \textbf{audience size} as the average number of live viewers the stream presented daily throughout our sampling period. Audience size ranged from 0 to 31,000 average concurrent viewers. We use standard clustering techniques to group streams with similar sized audiences, specifically using a mean shift algorithm to group streams with similar sized audiences. We opted to use a mean shift algorithm because it is based on a non-parametric density estimation, and therefore we would not need to know the number of clusters beforehand (unlike K-means). Our clustering algorithm identified five different audience scales, i.e., clusters.

\subsection{Modeling Audience Participation}

To allow a more nuanced understanding of our data, we selected a subset of streams from each cluster for further analysis. We used cluster sampling with the probability proportionate to size and based on the number of streams in each cluster. For the streams sampled from each cluster, we then modeled the participation of their audiences. Note that we focus our analysis on ``participatory audiences'', i.e., audience members who posted at least once in the stream's chat. 

To start to unravel audience participation, for each stream, we study the number of days that an audience member kept active continuously (i.e., in one day posted at least once in the chat). For each stream, we also model the number of chat messages sent by each audience member. To understand these messages in greater depth, we study the \textit{nature} of the chat messages posted in each stream. We build on Van Dijk and Goffman's work  \cite{van1997discourse,levinson1988putting}, suggesting that audience participation can be understood through textual analysis. 

Similar to \cite{van1997discourse}, our textual analysis studies usage of slang, pronouns, and ``salutations'' (i.e., phrases where people greet or say goodbye to each other). Together, these terms can act as a proxy to understand community building and relationship closeness between audiences and streamers in each stream \cite{ribak2002ask}. To identify slang, we use a combination of SlangNet \cite{dhuliawala2016slangnet} and manually captured Twitch specific slang over all chat messages. Previous work \cite{Ford2017Chat} also used similar techniques to identify slang in Twitch chat. To identify the amount of pronouns and salutations used we utilized pre-determined dictionaries, \footnote{\url{http://www.esldesk.com/vocabulary/pronouns}, \url{https://www.fluentu.com/blog/english/english-greetings-expressions/}, \url{https://www.ego4u.com/en/cram-up/vocabulary/people}} a technique other prior audience research \cite{savage2015}.


\subsection{Investigating  Streamers' Performance}
We use qualitative methods to examine streamers' performance and how audiences interact in practice to those performances. We examine stream video and chat data from a subset of streams using an approach informed by ethnomethodology and conversation analysis  \cite{tekin2017ways,pelikan2016nao}). 
In particular we were interested in the kinds of strategies that streamers adopted to trigger participation, and to better understand what types of interactions those strategies (performance) produced. We focused our analysis on unpacking and making  concrete the notion of streamers as ``performing'' for an audience. 


\section{Audience Participation and Streamer Performance at  Scale}

 Through large-scale data analysis based on audience size, we identified five clusters of channels on Twitch. Table \ref{table:Clusters} shows basic ``demographics'' of these streams and notes the data collected for each, while Table \ref{table:Results} shows characteristics derived from this data.

\begin{table*}[htp]
  \centering \small
  \begin{tabular}{p{3cm} p{2.0cm} p{2.0cm} p{2.5cm} p{2.7cm}}
   { \bf Cluster} & {\bf Avg Live Stream Viewers} & {\bf Audience Members in Chat}  & {\bf Total Audience Messages Collected}& {\bf Total Minutes of Video Stream Analyzed}\\
    \hline
{ 1: Clique} & 0-6 & 1,374 & 49,909 & 438 \\
{ 2: Rising Streamers} & 6 - 1,879 & 54,526 & 723,928 & 1,071 \\
{ 3: ChatterBox} & 1,879 - 7,703 & 169,546 & 3,166,399 & 506\\
{ 4: Spotlight Streamers}& 7,703 - 21,678 & 329,279 & 3,925,338 & 881\\
{ 5: Celebrities and Tournaments}& 21,678+ & 189,737  & 3,884,273 & 1,101\\ \hline

  \end{tabular}
  \caption{Overview of the characteristics of each cluster}
  \label{table:Clusters}
\end{table*}

\begin{table*}[htp]
  \centering \small
  \begin{tabular}{p{3cm} p{1.5cm} p{1.5cm} p{1.5cm} p{1.5cm} p{1.5cm} p{1.5cm} p{1.5cm}}
   { \bf Cluster} & {\bf Avg Live Stream Viewers} & {\bf Avg Number of Bots}  & {\bf \% Positive Msg }& {\bf \% Negative Msg} & {\bf \% Exit Rate on Day 1} & \bf{Avg Words per Message}\\
    \hline
{ 1: Clique} & 0-6 & 1.25 & 23.98 & 9.86 &  65.21 & 6\\
{ 2: Rising Streamers} & 6 - 1,879 & 1.57 & 21.25 & 10.15 &  64.74 & 5.71\\
{ 3: ChatterBox} & 1,879 - 7,703 & 2.12 & 14.38 & 7.44 &  56.51 & 4.36\\
{ 4: Spotlight Streamers}& 7,703 - 21,678 & 1.87 & 11.27 & 6.05 & 58.06 & 3.82\\
{ 5: Celebrities and Tournaments}& 21,678+ & 1.5  & 13.97 & 7.17 & 51.49 & 3.99\\ \hline

  \end{tabular}
  \caption{Overview of the resulting characteristics of each cluster}
  \label{table:Results}
\end{table*}

\subsection{Cluster 1 (Clique Streams):} 

\subsubsection{Audience Size and Messages.} 
Channels in this cluster had small audiences, ranging from 0 to 6 live viewers on average. This cluster is characterized by having the longest messages of all clusters, with 6 words per message on average.

Audience members in this cluster produced an average of 36 messages per user. This was the highest average number of messages produced. However, this allowed the discussion to be dominated by a small cadre of viewers, perhaps driving away new potential community members.

To further explore this dynamic, we examined the channel chat using TF-IDF (\textit{Term frequency - Inverse document frequency}). Some of the most frequently used words in this cluster were related to life outside of video game streaming, and to existing friendships. The words most commonly used by audiences in this cluster were: lol, play, like, good, hi, game; and frequent phrases included  \emph{``Are you going to play [video game] when it comes out?''}. We identified that streamers usually talked about the games they were playing, but also about their personal life, and even their work. These behaviours were replicated in chat by audience members. 

Another hint at the relationship-driven nature of these channels is that the audience notifies the streamer when leaving a day's stream and, in some cases, even provided the reason for their exit (3.15\% of the audience members said goodbye when leaving the stream v.s. 0.33\% in all other clusters). At the same time, we observed that, in this cluster, audiences mentioned each other the least in the chat (2.16\% of the messages had mentions towards other audience members v.s. the median of 8.79\% in all other clusters). This behavior likely happens because, given that it is a relatively small group of people who are present, there is no need to use mentions to identify who is being addressed. 

Streamers, on the other hand, were likely to mention audience members, particularly newcomers. When newcomers arrived to the stream and started participating in the chat, the streamer and other audience members would usually call them out publicly and welcome them to the stream. We saw phrases such as: \emph{``Hey <username>, welcome to the stream.''}. 13.59\% of all mentions in this cluster were targeted towards newcomers. We hypothesize that streamers were attempting to create a stronger bond with them, similar to the behavior that is observed on Facebook when individuals are tagged or mentioned in posts \cite{savage2015tag}. 

\subsubsection{Audience's Number of Days Active.} Among our five clusters, this cluster had the most difficulty with retention. 65\% of the audience that participates in chat stays active for only one day, and 90\% of visitors appear in chat on only two or fewer days, a very low retention rate. 


Getting visibility on Twitch for one's stream, along with keeping the existing audience active, seems to be one of the main problems that streams of this cluster suffered from the most. The reason why keeping the audience active  might be difficult is that most streamers had sporadic streaming schedules. This sporadic aspect of their interaction does not seem to  allow audience members to plan or schedule time to see the stream, and hence they stop participating. Separately our results also showcased that most streamers typically do not stay in Cluster 1 for more than 15 days. They either drop out or are able to recruit a large enough audience to now be considered part of Cluster 2. This suggests that streamers sitting at this audience size would be prime targets for support or supportive technologies.

We interpret the above data to mean that this group is primarily composed of streamers who are new to streaming. It also appeared that the audiences that participated with the streamer were close friends with the streamer, not only online but also involving shared offline activities. Together this behavior lead us to name this cluster ``Clique Streams.''

\begin{figure}[!h]
    \centering
    \includegraphics[scale=0.5]{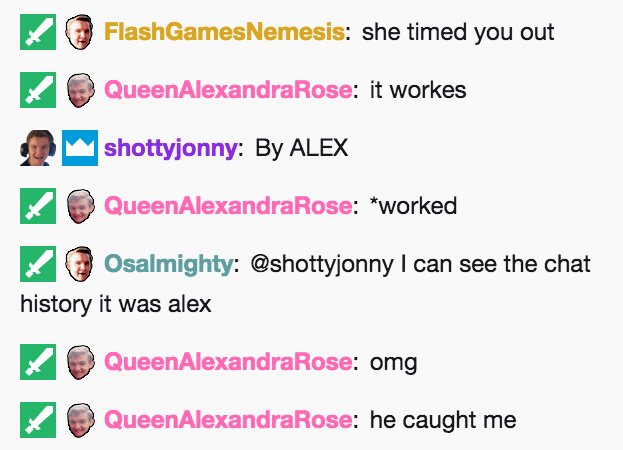}
    \caption{Moderators (users with green swords in chat) participating in chat on Cluster 2}
    \label{fig:moderators}
\end{figure}

\subsection{Cluster 2 (Rising Streamers):} 

\subsubsection{Audience Size and Messages.}
Despite high turnover, the streams in this cluster presented a consistent and relatively large audience size: the average number of audience members per stream was 339 (see also Table \ref{table:Clusters}). 

These streams also start a formal integration of bots, in particular to facilitate moderation. Bots release moderators from a heavy workload by automating some moderation functions, particularly when chat is rapid. Bots were also used in these streams to  greet newcomers, answer questions, promote the streamer's social media accounts and interact with viewers. An interesting aspect of these streamers is that although they did not have the largest audiences, they averaged 1.57 bots in their stream, more than the small Clique Streams (1.25) and as many as the large-scale Celebrities and Tournaments (1.5).

\subsubsection{Audience's Number of Days Active.}
This cluster presents a larger participatory audience than Clique Streams, but streamers in this cluster still struggle with retention. We observed that 64\% of the audience participates in chat for only one day and an additional 19\% of the audience participates for a maximum of two days. The average number of days an audience member was active in this cluster was 2, compared to 2.31 for Clique Streams. 

\subsubsection{Streamer Performance.}
Streamers in this clusters were typically invested in professional-quality streaming equipment. For instance, a vast majority of the video streams we analyzed had streamers using ``green screens'' and multiple monitors for their live broadcasts. 

In addition to gaming streams, we observed that most non-gaming channels such as in-real-life (IRL), Talk show, and Creative streams fell into this cluster. It also included streamers who built community around shared identity or experience (e.g. transgender streamers), or who tended to cover specific topics in their stream (e.g. streamers who stuck to a specific game).

However, despite the use of a consistent schedule, the integration of bots, and the investment in streaming equipment, these streamers still struggled in recruiting large audiences (most streamers in this cluster had generally less than 500 audience members). For some streams, it may be because they are \textit{successfully} directing themselves to a smaller, specific audience rather than \textit{unsuccessfully} recruiting a large and general one. However, other streamers may be bidding for a more general audience and simply cannot retain users long enough to manage it. Together this behavior lead us to name this cluster ``The Rising Streamers.''

We also note that streams that belong to this cluster can apply to the ``Affiliate Program,'' or Twitch partnership in which streamers can start to receive subscriptions to their channel and are granted custom chat emoticons meaning that streamers can begin to monetize their streams.


\subsection{Cluster 3 (The ChatterBoxes):} 

\subsubsection{Audience Size and Messages.} This cluster has from 1,879 to 7,703 active audience members (See Table \ref{table:Clusters}). Audience members in this cluster of streams were the second-most participatory of all (after Cluster 1). Audience members also referenced each other the most using the \textit{"@username"} chat structure, likely hinting that they were still having conversations with each other but the increasing volume of chat messages required them to formally seek each other's attention.

This increase in the velocity in which messages are sent in chat was substantial. Streams in this cluster presented a median speed of 1.95 messages per second in comparison to less than 0.00067 for Clique Streams and 0.57 for Rising Streamers.

Another interesting fact about this cluster is that the use of slang and stream-personalized emoticons starts becomes present in chat, meaning that in this cluster the audience might be building a sense of community and identity around the stream. This is of particular interest given that stream-personalized emoticons are \textit{available} to Rising Streamers, but not typically adopted. 

\subsubsection{Audience's Number of Days Active.} This cluster demonstrates a massive increase in ability to retain audiences compared to the Rising Streamers and Clique Streams. Compared to previous clusters, the exit rate on day one decreases dramatically, from 64\% to 56\%, meaning that streamers here were able to hold their audiences' attention. Streamers are able to maintain a significant audience size, with an average of over 4,468 individuals per stream. Audience members in this cluster stayed active for an average of 2.34 days. 

\subsubsection{Streamers' Performance.}
In this cluster, two individuals were typically present in the stream. For example, we observed channels with one streamer playing the game live, while another reads aloud the messages from the audience and initiates discussions with the audience based on what they typed. This dynamic facilitates having lengthy conversations with a reasonably large audience.


Given that this cluster appeared to engage in the most lengthy discussions, and actively practiced mentioning and engaging in conversations with other audience members, we named this cluster ``The ChatterBoxes''. It is noteworthy that large audiences were able to have more extended conversations than smaller audiences; it suggests that bottlenecks on streamer attention are more easily replaced with additional chat participation than with technology alone.

\subsection{Cluster 4 (Spotlight Streamers):}

\subsubsection{Audience Size and Messages.} This cluster has a massive number of concurrent viewers, ranging from 7,703 to 21,678 live viewers on average (See Table \ref{table:Clusters}). This cluster has the characteristic that its streams presented the largest proportion of the participatory audience on Twitch. The total number of active audience members was 329,000, double the size of cluster 5.

\textit{Audience's Number of Days Active.} We analyzed how streamers held this large scale participatory audience through time and observed that even though streamers temporarily captured a massive audience, 58\% of the audience members in these channels would only participate in chat for one day. However, we did not observe ``raid type'' departures in our samples. Audiences arrive sporadically to the stream and leave sporadically; it did not seem to be a coordinated effort. Compared to the neighbor clusters 3 and 5, this cluster has the highest percentage of members that left after participating for only one day. The messages that audience members sent in were also the shortest of all clusters (a median of 3.82 words in comparison to 5.01 for other clusters).

\subsubsection{Streamers' Performance.}
While streamers in this cluster did address and react to their audience, the reactions were usually not targeted towards particular individuals. For instance, streamers in this cluster rarely mentioned to whom they were responding to. Streamers of this cluster appeared to focus more on talking out loud about their strategies and plans than necessarily engaging in conversations with their audience. This behavior might explain why this cluster showcased the least participatory audience members:  streamers did not seem to promote or encourage conversations.

Streamers in this cluster talked the most about their struggle with playing the video game and interacting with the audience. For instance, it was common to hear streamers state that they could not ``aim, jump, shoot and talk at the same time.'' We also observed streamers in this cluster fall silent while playing, only speaking when something went wrong in the game. In some cases, streamers tended to spend the majority of their time being quiet and focused while playing, especially in moments of tension. Hence these streamers likely had to focus all their energies on playing the game rather than interacting with the audience. 

All the streams in this cluster were highly promoted by Twitch and had large audiences but retained relatively few of their new viewers. This trajectory led to us naming this cluster ``The Spotlight Streamers.''  We believe the process of becoming a spotlight streamer might not have prepared streamers adequately to deal with the influx of viewers when they were highlighted on the front page of Twitch, and consequently viewers would lose interest and leave after relatively short visits. It also seems that being a spotlight streamer might have been very stressful for streamers, and hence they had to limit interactions with their audience to focus on the game. This dynamic of having to be silent for concentration likely caused audiences to leave as there were limited engagement opportunities for them. 

\begin{figure}
    \centering
    \includegraphics[scale=0.4]{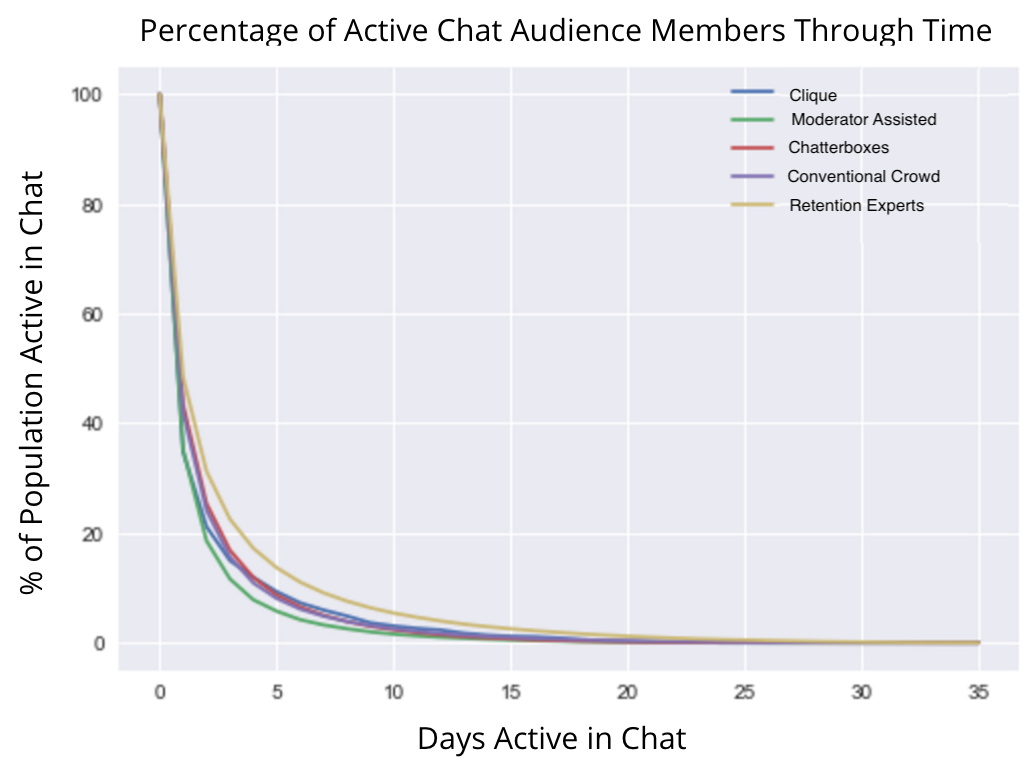}
    \caption{Chat retention over time.}
    \label{fig:retention}
\end{figure}

\subsection{Cluster 5 (Celebrities and Tournaments):}

\subsubsection{Audience's Number of Days Active.}
 This cluster included the most popular streams on Twitch, which also held the largest audience for a longer term. Streams in this cluster retain on average over 1,000 audience members for more than 20 days per stream. Overall, streamers in this cluster seemed to be effective at entertaining their audiences.
 
 It is also important to mention that the most popular tournament streams also fall into this category. We're able to see that this type of streams receive massive audiences during the short period of time that the tournaments was streamed. Tournaments that had live audiences involved invest a considerable amount of resources on replicating the logistics, look and feel of other sports in a sense that they rent a space to hold the tournament, sometimes even stadiums, decorate the venue and hire hosts to narrate the matches.

\subsubsection{Audience Size and Messages.}
Streams in this cluster presented enormous live audience sizes (21,678+ participatory audience members). However the number of words used in each chat message was very small, with an average of 4 words (see Table \ref{table:Results}). This short text exchange might have been due to the crowd/stadium environment that appeared to emerge in these streams. The message exchange speed here was 2.45 messages per second, almost three times faster than Cluster 4. Also, messages on this cluster were highly likely to contain custom emoticons, which allow the viewer to make visible their feelings or mood about the current situation of the stream without getting lost in the rapidly scrolling chat. This rapid interchange of low-content messages creates a sense of ``crowdspeak'', where audiences are communicating similar to that of a stadium/crowd environment \cite{Ford2017Chat}. 

\subsubsection{Streamers' Performance.}
Streamers in this cluster were typically professional Twitch streamers with verified accounts, and were among the top players of a particular  game. These streamers were the most consistent in their live transmissions: their schedule was posted on their profile, and in our data analysis we identified that during the study period they never missed a day of streaming. When the streamer leaves the stream, even for a short break to get food or use the bathroom, they lose on average 5\% of their viewers. They also stream for long periods of time (4 hours or more in a session) and typically play one of the five most popular games on Twitch. 

These streamers generally provided entertaining reactions to audience's participation, such as interweaving jokes or criticizing opposing players' skill at the game. However, the interactions the streamer has with their chat decreases while they play with other people on stream. 

On tournament streams, they always presented a host whose job was to narrate the actions the players were doing on the match in a similar tone that other hosts would narrate other sports. They also showed a balanced ratio between talking on stream and keeping silent sections of it giving a hint that they are professional hosts that have broad knowledge on how to manage and communicate to the audience. This streams implement multiple sections of the stream that include, interviews with the players, short brakes between matches, game analytic, discussions and sponsored content.

Due to the high number of viewers in the stream, managing notifications and responding to subscriptions becomes difficult for the streamer due to the high velocity of subscriptions. In order to solve this, streamers dedicate spaces of time where they don't need to concentrate deeply, like game loading sections or lobbies, to read out loud the subscriptions they received during the time they were playing. Donations were structured as a way to get the attention of the streamer, since a financial contribution made their message visible for the streamer to interact with it. This behavior led us to name this cluster the ``Celebrities and Tournaments.''




\section{Understanding Aspects of Streamers' Performance}

Throughout our qualitative analysis over the video streams and chat messages on Twitch, we identified key practical problems that streamers face when mobilizing participation of their audience in both small and massive chats:

\begin{itemize}
    \item Interweaving attention to the participation of their audience (e.g. glances at the chat stream) with attention to the game.
    \item Selecting which participatory elements (e.g. chat messages or subscriptions) to react to, especially connecting their reaction back to the specific element that prompted it.
    \item Disambiguation of their own reactions to chat messages (e.g., pairing reactions to chat messages or other events such as subscriptions).
    \item Performing the above with an appropriate frequency to the volume and content of chat messages.
\end{itemize}


While the sources of streamer-audience interactions include chat messages, subscriptions, and donations, we observed that subscriptions and donations do not seem to typically lead to extended interactions. Instead, financial transactions seem to offer a more constrained possible reaction ``space'' for streamers - i.e. responding with a recognition of the audience action coupled with a positive assessment of some form (``thanks for the subscription!''). As part of this, streamers seem much more likely to incorporate identifying elements (i.e. usernames) into the recognition of subscriptions and donations than they are for chat messages. This practice does not seem driven by the need for disambiguation (i.e. clearly identifying one audience member from another) since subscriptions and donations are delivered directly to the stream video and not through the chat stream. Instead, it might serve more to motivate other audience members to carry out similar actions \cite{banerjee1992simple}. 

When it comes to chat, one of the primary methods by which streamers draw on audience generated content is via potentially paired ``reading'' and ``reaction'' practices. By reading practices, we mean to say that streamers are reading either complete chat messages or fragments of them out loud to the audience as they play. Rarely is any other identifying element provided alongside this (e.g., no usernames). Typically readings of this form are complete reproduction ``verbatim'' utterances which may be produced at ``any'' moment (often e.g., bringing to a close some existing ``reaction'' that the audience generated). Sometimes such readings are marked sporadically as well, e.g. by transforming a chat comment into a question. 

Reactions are when a streamer does something that is seen as paired with a chat message. For instance, this might include answering a question or carrying out a suggestion in-game. Reactions encompass basic question-answer pairs through to extended monologues by the streamer (e.g., telling a story). It is important to note that reactions are not necessarily explicitly paired with any reading-out-loud. This perhaps indicates how strong the assumed shared perspective mentioned above actually is between streamer and audience. It is unclear so far in our analysis why some reactions are paired with readings while others are not - i.e. the disambiguation problem for streamers seems to be partial. 

The generally terse quality of streamers doing reading out loud seems remarkable given the number of messages appearing on screen, and the speed at which they disappear. However, reading practices build upon an assumed shared or partly symmetrical perspective between streamer and audience in terms of what is happening in\-game, what a streamer might visibly be doing on camera (assuming they are visible at all), and what chat messages are visible currently. All participants seem to assume that there is a reasonably tight temporality between the production of the chat message and it being ``taken up'' by the streamer in some way.

Another key method of interaction with audience is for streamers to ``request'' something. This often involves obtaining an answer to a question (game-relevant or not) or a request for a resource, such as a music recommendation and/or a link to music to listen to or play. Requests take on an organization that is sensitive to the volume of chat messages. For streamers with a larger audience and higher velocity of chat messages, some streamers we observed formulated their requests with embedded ``voting'' options as possible responses for audience, for example where the streamer asks the audience what in-game action they should perform, listing options and then requesting a response of e.g., ``1 or 2''.

Our research to date explores audience actions as sources for streamer action, but we expect that they may at times lead to longer strings of interaction. We noted that chat messages in particular (although also occasionally true of messages connected with donations and subscriptions) can function as ``topic continuers'', i.e. leading to streamers performing a connected sequence of reading-reaction that takes its sense in some way from a prior pair.

\section{Discussion}

Our large scale analysis shows that streamers perform differently when they interact with audiences of differing sizes. For instance, when interacting with large audiences, some  streamers engaged in ``self-mockery,'' whereas with small audiences streamers usually presented themselves in a positive light. We also observed that audiences participate differently across audience sizes, i.e. not all clusters are equally participatory or in the same way. For instance, the audiences in Cluster 1 (Clique Streams) presented the most intimate/relational messages (e.g. audience members frequently said goodbye to each other), while audiences in Cluster 3 (The ChatterBoxes) were the most communicative, posting the most number of messages and the longest. 

We found that once a streamer gets more than six regular average concurrent viewers, the role of the moderator becomes much more important; while moderators are known to be crucial in large Twitch streams, it is surprising how important they seem to be in even relatively small streams.

In the following, we discuss our results in terms of the design implications for the different stakeholders involved in audience participation games: a) streamers, b) game designers, and c) platform developers. 

\subsection{Implications for Streamers}
Our results showed that streamer attention is split between playing the game, and playing to the audience. Another challenge that streamers face is that audience sizes change over time. For example, we hypothesize that our Cluster 4 streamers (Spotlight Streamers Streams) had trouble retaining audience participation over time, because they may have had sudden or sporadic audience growth. Therefore, these streamers were dealing with a larger audience than what they (streamers) were  used to managing. Meanwhile, new audience members would be unfamiliar with the stream's chat culture, making it difficult to interact in a way that would be likely to be received positively. 

Streamers do not have ways to practice retaining audiences at new scales before they are asked to do so in a live situation. Designers can respond to these challenges by creating tools that help streamers accomplish their goals - and by making those tools audience-size-sensitive, since streamer goals, and the best ways of accomplishing such goals, will vary by scale.

One first step might be to create visualization tools that in real-time can show streamers the results from our audience analysis. For example, the tool could show when a large number of newcomers join the stream, the uptake of custom emoji as a proxy for success of their brand, or provide real-time text mining in order to help streamers understand what their chat is talking about. Here it might help to adopt some of the tools and techniques that previous research and practitioners have created for visualizing and interacting with online audiences, such as providing summaries of online discussions \cite{welser2007visualizing} or showcasing the areas of expertise of audience members \cite{savage2014visualizing}.

Given the different ways that different sized audiences participated with streamers, it might also be helpful for designers to consider creating tools that can detect streamer behaviors (e.g. reading chat messages out loud) and identify the best practices (i.e., behaviors) that are most likely to help foster engagement. 

Broadly, we propose that incorporating metrics related to  audience size as a core design factor in both data tools for streamers, and research and analysis tools for systematizing best practices of streamer behavior.
 
\subsection{Implications for Game Designers}
Through our study, we identified that many streamers struggled to perform for their audience while effectively playing their game. For example, in streams with large chats, it was fairly complicated to maintain a conversation while messages rapidly scrolled by. It can therefore be important for game designers to envision tools that help streamers to effectively manage all of these elements: to maintain the ``persona'' or image they want to convey for their audience, keep their audience engaged, and still adequately and effectively play their game. 

Designers might use phase-based design strategies \cite{bergstrom2010exploring} to allow both streamers and audiences to participate differently during different phases of play. For example, existing collectible card games typically have a slow, reflective phase of deck-building, which can be done solo and has no time limit; actual matches are competitive, often tense, and typically include time limits on play \cite{paul2011optimizing}. During slower, reflective gameplay phases, streamers can focus on connecting with their audiences; during time-sensitive or multiplayer phases, they can focus on entertaining their audiences with striking moments of gameplay. These dynamics exist in streams focused around these games, but similar mechanics could be valuable in other types of games. 

Designers might also consider how to make audience interactions \textit{part} of gameplay \cite{seering2017APG}. For example, for large audiences, one could imagine a ``Where's Waldo'' game type approach, where the streamer has to identify certain types of interactions or individuals in her audience \cite{handford1987s}. 

Finally, our analysis also identified that streams had substantially different audience dropout rates. For instance, in Cluster 1 (Clique Streams) where streamers lacked a formal schedule, audiences usually came and went into the stream at different times. In other words, audience members were likely not to be present for the whole game. When designing games for streaming, designers must consider how to onboard audiences into gameplay at any point. This might take different forms. Game rounds could be short and independent, to quickly put newcomers on the same footing as everyone else; the game itself could contain all necessary information to interpret game state and trajectory; new viewers might receive special onboarding information from bots, moderators, or other community members; or games could provide unique interfaces for new viewers using the Twitch API. Designers should also consider whether the design of these onboarding interfaces could communicate social norms about the game, for example whether it is socially appropriate to engage in personal conversations or to poke fun at the streamer.

\subsection{Implications for Platform Designers}

In our study, we identified that some streamers lost a large number of their audience members over relatively short periods of time. It can, therefore, be useful for platform owners and designers to think about providing training or onboarding help for streamers, perhaps exploring in collaboration with researchers to study which stream behaviors or tools are most effective at retaining viewers. This especially can provide value for platform owners as it can facilitate engagement and retention on their sites and systems.

Best practices could also be embedded directly in interfaces that adapt as audience sizes change; for example, the traditional chatbox could be designed to look and function significantly differently for different sizes of streams. The relatively new ``rooms'' feature facilitates some of this, but much more could be done. Twitch, or even extension designers, could create tools that summarize or provide meta-commentary for on-stream actions. Note that Twitch does currently provide some aid to streamers by providing metrics for Associate/Partnered streamers \cite{twitchDinero}. However, streamers could make use of improved \textit{strategies} and \textit{tools} for engaging their audience.

As we discuss above, most streamers in Cluster 1 typically do not stay in this cluster for more than 15 days either growing into Cluster 2 or dropping out entirely. Therefore, it might be important for platform designers to think about how to provide support to help retain streamers during that critical period, in order to empower streamers to ``rise up'' instead of drop out. This group is a prime target for research identifying what behaviors and tools enable some streamers to grow and cause others to drop-out.

While we frame audiences' participation here broadly as binary, either participating in a channel at some rate or not participating at all, our results suggest that there might be value in thinking now about interaction where audiences can appear at ``just the right time'' to act as viewers or as participatory audience members. For example, there might be times when streamers or platform designers would want large audiences to be present in order to view special moments that are taking place within the platform, such as the breaking of a speedrun record. Though not an easy technical or design challenge, both Twitch and streamers would benefit from tools that were able to identify highlights with just enough time before they happen for viewers to tune in to experience them. 

Beyond identification of these moments, another design question here is thus around how to design interfaces that best mobilize audiences to participate in certain streams at certain times. It could be very helpful here to build from previous research on mobilizing strangers online to take small micro actions \cite{savage2016botivist,nichols2012asking}.

Finally, for tool designers, it can also be quite helpful to have systems for reflecting ``stream culture,'' because tools that better match the needs of the stream, e.g. moderation tools that are sensitive to appropriate emoticons, are much more likely to be both adopted and successfully used. Facilitating such types of tools might be especially important for many different types of users, e.g. LGBT streamers talking frankly about sex education and gender. Within this design space, we could also envision hybrid systems that mix automated methods with human judgments (e.g. judgments from moderators about what is appropriate and inappropriate for a stream) to learn, through time, the norms within different streams and teach them to viewers, particularly those who are new to the space.

\section{Limitations and Future Work}


The insights that this work provide are limited by the methodology and population with which we studied. For example, Twitch may attract a specialized subset of streamers and viewers, hence our results may not generalize to all other online streaming platforms. Future work could focus on studying a wider spread of live streaming platforms, in order to better understand this phenomenon.

Also, the methods we used focused on breath rather than on depth. Future work could be seen though in-depth participatory or even contextual interviews with streamers and their audience members. Our analysis also focused primarily on participatory audiences and did not investigate lurkers (i.e., audiences that simply viewed the stream). Future work can investigate the dynamics between participatory audiences, lurkers, and streamer interactions. Here, research could build off previous work that has investigated how people's behavior changes depending on how they image their lurkers \cite{bernstein2013quantifying}. 

In the data we collected we also did not see any evidence of substantial raids (i.e., large number of audience members joining a particular stream at the same time as directed by another streamer). Future work could investigate how raiding affects audience participation in both the short and long term. Note also that for Clusters 1-3, we did not find significant difference between Twitch partnered channels or affiliates and channels that were not partnered or affiliated. None used a subscriber only mode for chat. All channels in cluster 4 and 5 were partnered, so we cannot make comparisons for these size groups. Future work however could explore in greater depth how having a formal partnership with platforms, such as Twitch, transforms the interactions between streamers and audiences. 
\section{Conclusion}

In this paper, we investigate Twitch as a vehicle for understanding audience participation and streamer performance at different scales within online live streaming platforms. Our research shows the characteristics presented by streamers and audiences on five different audience size scales: Clique Streams, Rising Streamers, The Chatterboxes, Spotlight Streamers and Celebrities and Tournaments. 

We present tools and design implications for streamers, game designers and platforms designers to address various obstacles observed for these clusters that ultimately could improve the experience of streamers and audiences.

%
\bibliographystyle{ACM-Reference-Format}
\bibliography{Twitch}

%
\typeout{get arXiv to do 4 passes: Label(s) may have changed. Rerun}

\end{document}